\documentclass[article,nofootinbib]{revtex4-1}
\usepackage{subfigure}
\usepackage{float}
\usepackage{epsfig}
\usepackage{dsfont}
\usepackage{amsmath}
\usepackage{braket}
\usepackage{amsfonts}
\usepackage{amssymb}
\usepackage{graphicx}  
\usepackage[dvipsnames]{xcolor}
\usepackage{todonotes}
\usepackage{bbm}
\usepackage{tikz}
\usepackage{lmodern}
\usetikzlibrary{decorations.pathmorphing}
\tikzset{snake it/.style={decorate, decoration=snake}}
\expandafter\ifx\csname package@font\endcsname\relax\else
\expandafter\expandafter
\expandafter\usepackage
\expandafter\expandafter
\expandafter{\csname package@font\endcsname}%
\fi

\RequirePackage[colorlinks,citecolor=blue,urlcolor=magenta,linkcolor=blue]{hyperref}

\newcommand{\be}{\begin{equation}}
\newcommand{\ee}{\end{equation}}
\newcommand{\bea}{\begin{eqnarray}}
\newcommand{\eea}{\end{eqnarray}}

\begin{document}
\title{Aspects of gravitational decoherence in neutrino lensing}
\author{Himanshu Swami}
\email{himanshuswami@iisermohali.ac.in}
\author{Kinjalk Lochan}
\email{kinjalk@iisermohali.ac.in}
\affiliation{Department of Physical Sciences, Indian Institute of Science Education \& Research (IISER) Mohali, Sector 81 SAS Nagar, Manauli PO 140306 Punjab India.}
\author{Ketan M. Patel}
\email{kmpatel@prl.res.in}
\affiliation{Theoretical Physics Division, Physical Research Laboratory, Navarangpura, Ahmedabad-380 009, India.}

\begin{abstract}
We study decoherence effects in neutrino flavour oscillations in curved spacetime with particular  emphasis on the lensing in a Schwarzschild geometry. Assuming Gaussian wave packets for neutrinos, we argue that the decoherence length derived from the exponential suppression of the flavour transition amplitude depends on the proper time of the geodesic connecting the events of the production and detection in general gravitational setting. In the weak gravity limit, the proper time between two events of given proper distance is smaller than that in the flat spacetime. Therefore, in presence of a Schwarzschild object, the neutrino wave packets have to travel relatively more physical distance in space to lapse the same amount of proper time before they decoher. For non-radial propagation applicable to the lensing phenomena, we show that the  decoherence, in general, is sensitive to the absolute values of neutrino masses as well as the classical trajectories taken by neutrinos between the source and detector along with the spatial widths of neutrino wave packets. At distances beyond the decoherence length, the probability of neutrino flavour transition due to lensing attains a value which depends only on the leptonic mixing parameters. Hence, the observability of neutrino lensing significantly depends on these parameters and in-turn the lensing can provide useful information about them.
\end{abstract}

\maketitle
	
\section{Introduction}
Neutrino oscillations, in curved spacetime has gained attention in recent times \cite{Cardall:1996cd,Fornengo:1996ef}, for the reason that such analysis is not only sensitive to the background geometry and hence the gravity theory at work \cite{Ahluwalia:1996ev,Ahluwalia:1996wb,Grossman:1996eh,Bhattacharya:1996xb,Luongo:2011zza,Geralico:2012zt, Koutsoumbas:2019fkn, Buoninfante:2019der, Blasone:2019jtj} but they also reveal some salient features of the neutrino sector which are not present in flat spacetime \cite{Sorge:2007zza, Lambiase:2005gt}. Apart from increasing  the oscillation length of neutrinos, effects such as spin-flip or helicity transitions \cite{Sorge:2007zza,Lambiase:2005gt} and possible violation of equivalence principle \cite{Lambiase:2001jr,Bhattacharya:1999na} have been investigated in the gravitational settings. Gravitational lensing where different trajectories of neutrinos  around a massive astrophysical body get focused on a point of detection and its effects on flavour oscillations are studied in the context of Schwarzschild geometry in \cite{Fornengo:1996ef,Crocker:2003cw,Alexandre:2018crg,Dvornikov:2019fhi,Capolupo:2020wlx,Eiroa:2008ks}. In our previous work \cite{Swami:2020qdi}, we studied weak gravitational lensing of neutrino by a Schwarzschild mass and showed that the resulting flavour oscillations become sensitive to the individual masses of the neutrinos, paving a potential way of measurement of individual masses of different neutrino species.

Interesting as these results may appear, like in the flat space, the neutrino oscillations  in most of such considerations, have been studied using plane wave approximation. In a realistic generic scenario though a wave packet approach is more practical as the neutrinos are produced and detected as localised wave packets of finite width in position space. Introduction of wave packet in such studies, introduces a new length scale beyond which the neutrino oscillations cease \cite{Giunti:1991ca,Giunti:1997sk,Giunti:1997wq,Grimus:1998uh,Grimus:1999ra,Akhmedov:2009rb,Akhmedov:2010ms}. Owing to the non-trivial leptonic mixing, a wave packet of neutrino created in a particular flavor in a weak interaction process can be decomposed in terms of wave packets of different mass eigenstates. Under the time evolution these wave packets travel with different group velocities due to the different masses. Eventually, they get separated and the overlap between them drops off to an  insignificant value such that the probability of transition amongst the different flavours saturates to a value which depends only on the parameters of leptonic mixing, a phenomena known as {\it decoherence}. 

While the decoherence effects have been widely studied for neutrino oscillations in flat spacetime, it has attracted relatively less attention for neutrino propagation in the curved geometry. Recently, these effects have been investigated in \cite{Chatelain:2019nkf, Petruzziello:2020wea} for neutrinos travelling radially inward or outward in the background of the Schwarzschild metric. In this paper, we study the impact of decoherence on neutrino oscillation when the neutrinos get lensed by a gravitating object located in between the source and detector. It is seen that the spatial distances neutrinos cover before the onset of decoherence is larger than that for the flat spacetime. We find that, from a perspective of an observer at infinity, the decoherence coordinate distance does not explicitly depend on the mass of the gravitating body in case of the radial propagation of the neutrino wave packets while such a distance explicitly depends on the Schwarzschild mass in the case of the non-radial propagation. Further,  we observe that the decoherence in the gravitating scenario is sensitive to the individual masses of the neutrinos and not only on their squared mass differences. Therefore, monitoring of the decoherence provides an avenue for the mass estimates of the individual neutrino species contributing into the decoherence process.

In section \ref{DecoherenceFormalism}, we discuss the wave packet formalism for neutrino lensing and formulate the condition for decoherence. In section \ref{SchwarzschildDecoherence}, we use the formalism for the Schwarzschild geometry and discuss decoherence for the radial and non-radial propagation of the wave packets and compare it with the flat space case. We explicitly study non-radial propagation relevant for lensing scenario in section \ref{Lensing} and obtain the decoherence condition. Finally, we summarise the main results in section \ref{Summary}.

	
\section {Neutrino wave packet and decoherence in curved spacetime} \label{DecoherenceFormalism}
Consider a neutrino in flavour eigenstate, $\nu_\alpha$, produced in some weak interaction process occurring during a spacetime interval centered at the source coordinate $(t,\vec{x}) = (t_S, \vec{x}_S)$.  The state can be expressed in terms of wave packet as \cite{Akhmedov:2009rb}
\be \label{stateS}
|\nu_\alpha (t, \vec{x}) \rangle = \sum_i\, U_{\alpha i}^*\, \psi^S_i(t,\vec{x})\, |\nu_i\rangle\,, 
\ee
where $U$ is the lepton mixing matrix and index $i$ corresponds to the neutrino mass eigenstate. In flat spacetime, a wave packet can be expanded unambiguously in the momentum basis. This advantage is somewhat lost in curved spacetime because the definition of the  momentum depends on the location of observer. Nevertheless, it is possible to define local Fourier transform by using a non-coordinate basis following the tetrad formalism \cite{Mitsou:2019nhj}. This allows one to write a spacetime evolved wave packet (at $t > t_S$) as
\begin{equation} \label{psiS}
\psi_i^S(t,\vec{X}(x)) = \int \frac{d^3p}{(2\pi)^3}\,f^S_{i}\left(\vec{p},\vec{p}^S_i\right)\, e^{i p_a X^a}\, e^{-i \Phi^m_i}\,,
\end{equation}
where $p_a $ and $X^a$ are the momentum and  position co-ordinates  in the tetrad basis of the tangent space around each spacetime point $x^\mu$.  The parameter
$a$  runs from $1$ to $3$, and $f^S_{i}\left(\vec{p},\vec{p}^S_i\right)$ is the momentum distribution function of neutrino produced at the source while $\vec{p}^S_i$ is the average momentum. The phase in the second exponent accounts for the propagation of the neutrino wave packet. In the curved spacetime, it is given by \cite{Cardall:1996cd}
\be \label{Phi}
\Phi^m_i = \int_S^D\, p_\mu^{(i)}\,dx^\mu\,,\ee
with $p^{(i)}_\mu = m_i g_{\mu \nu} dx^\nu/ds $ and $ds$ is the line element along the neutrino trajectory. Note that when there are more than one trajectories allowed in between the production and detection, the evolved phase in the Eikonal approximation depends on the particular path taken by the propogating neutrino mass eigenstate \cite{Swami:2020qdi}. Therefore, we denote this path dependency by a superscript $m$. Finally, the detected neutrino flavour state $\nu_\beta$ can be described by a wave packet centred at $\vec{x} = \vec{x}_D$ and therefore
\be \label{stateD}
|\nu_\beta (\vec{x}) \rangle = \sum_i\, U_{\beta i}^*\, \psi^D_i(t,\vec{x})\, |\nu_i\rangle\,, 
\ee
with
\begin{equation} \label{psiD}
\psi_i^D(\vec{X}(x)) = \int \frac{d^3p}{(2\pi)^3}\,f^D_{i, \vec{x}}\left(\vec{p},\vec{p}^D_i\right)\, e^{i p_a X^a}\,.
\end{equation}
Note that Eqs. (\ref{stateD},\ref{psiD}) do not explicitly depend on time as the process of detection is assumed to be time independent \cite{Akhmedov:2009rb}.

The amplitude of flavour transition after the neutrino has travelled from the source to the detector on a classical trajectory denoted by $m$ can then be obtained using Eqs. (\ref{stateS},\ref{stateD}) as \cite{Akhmedov:2009rb,Akhmedov:2010ms}
\be \label{Amp}
{\cal A}^m_{\alpha \beta} \equiv \langle \nu_\beta(\vec{x}_D) | \nu_\alpha(t,\vec{x})\rangle = \sum_i U_{\beta i} U^*_{\alpha i}\,\int \frac{d^3p}{(2\pi)^3} f_{i, \vec{x} }^{D*}\left(\vec{p},\vec{p}^D_i\right)  f_{i, \vec{x} }^S\left(\vec{p},\vec{p}^S_i\right) e^{-i \Phi^m_i}\,.\ee
The amplitude, in general, depends on the overlap of spacetime evolved momentum distribution functions which in turn depends on the neutrino trajectories. The probability of transition $\nu_\alpha \to \nu_\beta$ can be computed from the amplitude using
\be \label{P}
P_{\alpha \beta}= \frac{\left| \sum_m \mathcal{A}^m_{\alpha \beta}\right|^2}{\sum_{\beta}\left| \sum_m \mathcal{A}^m_{\alpha \beta}\right|^2}\,. \ee
In this way of deriving the transition probability, the normalization is enforced through the conservation of probability and it depends on the paths as noted earlier in \cite{Swami:2020qdi}.

Further simplification of amplitude can be achieved if the momentum distribution functions of the source and detector are known. Assuming that $f^S_{i,\vec{x}}$ is a function which has a sharp peak around $\vec{p}^S_i$, we substitute $\Phi_i^m$ with its series expansion at $\vec{p} = \vec{p}^S_i$:
\be \label{Phi_sub}
\Phi^m_i(\vec{p}) = \Phi^m_{i}(\vec{p}^S_i)\,+\, \vec{X}^m_i \cdot (\vec{p} - \vec{p}^S_i)\,+\,{\cal O}(p^2)\,,
\ee
with $\vec{X}^m_i = \partial_{\vec{p}}\, \Phi^m_i (\vec{p} = \vec{p}^S_i)$, evaluated with respect to the momentum description corresponding to any chosen point in the spacetime. This replacement in Eq. (\ref{Amp}) leads to
\be \label{Amp2}
{\cal A}^m_{\alpha \beta} =  \sum_i U_{\beta i} U^*_{\alpha i}\, e^{-i \Phi^m_i}\, \int \frac{d^3p}{(2\pi)^3} f_{i, \vec{x} }^{D*}\left(\vec{p},\vec{p}^D_i\right)  f_{i, \vec{x} }^S\left(\vec{p},\vec{p}^S_i\right) e^{-i \vec{X}^m_i \cdot (\vec{p}-\vec{p}^S_i)}\,,\ee
where now onwards $\Phi^m_i$ denotes the phase evaluated at $\vec{p} = \vec{p}^S_i$ and we do not show $\vec{p}$ dependence of $\Phi^m_i$ for simplicity.

We further assume that the momentum distribution functions at the source and detector are Gaussians. Explicitly,
\be \label{Gaussians}
f_{i,\vec{x}}^{S,D}(\vec{p},\vec{p}_i^{S,D})  = \left(\frac{2\sqrt{\pi}}{\sigma_{i\, S,D} }\right)^\frac{3}{2} e^{-\frac{\left(\vec{p}-\vec{p}_i^{S,D}\right)^2}{2 \sigma^{2}_{i\,S,D}}}\,,
\ee
such that
\be \label{gauss_norm}
\int \frac{d^3p}{\left(2 \pi\right)^3}  \left| f_{i}^{S,D}(\vec{p},\vec{p}_i^{S,D})\right|^2  = 1\,. \ee
In the following, we take equal mean momentum limit and take $\vec{p}_i^{D,S} = \vec{p}^{D,S}$ and $\sigma_{i\,D,S} = \sigma_{D,S}$ for simple estimations. Together with this, the substitution of Eq.(\ref{Gaussians}) in Eq.(\ref{Amp2}) after some straight-forward algebraic manipulations gives 
\be \label{Amp3}
 \mathcal{A}^m_{\alpha\beta} = 2\sqrt{2} \left(\frac{\sigma_D \sigma_S}{\sigma_D^2+\sigma_S^2}\right)^{3/2}\, \sum_{i}U_{\beta i}U_{\alpha i}^{*}\,  
 e^{-i \left(\Phi_i^m - \frac{\sigma_S^2}{\sigma_D^2+\sigma_S^2} \vec{X}_i^m \cdot \left(\vec{p}^D-\vec{p}^S\right) \right)}\,
  e^{- \frac{\sigma_D^2\sigma_S^2 \left|\vec{X}_i^m \right|^2 + \left(\vec{p}^D-\vec{p}^S\right)^2}{2\left(\sigma_D^2+\sigma_S^2\right)}}\,.
\ee
The first exponential gives rise to the neutrino oscillations while the second is responsible for the damping of the amplitude due to wave packet separation. Even if one chooses the mean local momentum of the detector wave packet to match exactly with that of the source, i.e. $\vec{p}_i^D = \vec{p}_i^S$, the amplitude damps eventually due to propagation as long as both of $\sigma_{S,D}$ are non-vanishing. Using Eq.(\ref{P}) and the amplitude obtained in Eq.(\ref{Amp3}), the probability is evaluated as
\be \label{P2}
P_{\alpha\beta} =\frac{\sum \limits_{i,j} U_{\beta i}^{*}U_{\alpha i} U_{\beta j}U_{\alpha j}^{*}\, \sum \limits_{m,n} e^{- i \Phi_{ij}^{mn}}\, e^{- {\bf X}^{mn}_{ij}}}{\sum \limits_{i} U_{\alpha i} U_{\alpha i}^{*} \sum \limits_{m,n} e^{- i\Phi_{ii}^{mn}} e^{- {\bf X}^{mn}_{ii}}}\,, \ee
where

\be \label{Phi2}
\Phi_{ij}^{mn} \equiv  \left(\Phi_i^{m} - \Phi_j^{n} \right) - \frac{\bar{\sigma}^2}{\sigma_D^2} \left(\vec{p}^D-\vec{p}^S\right) \cdot \left(\vec{X}_i^m-\vec{X}_j^n\right)\,, \ee
\be \label{X}
{\bf X}^{mn}_{ij} \equiv \frac{1}{2}	\bar{\sigma}^2 \left(|\vec{X}_i^m|^2+ |\vec{X}_j^n|^2\right)\,, \ee
and $\bar{\sigma}^2 = \sigma_D^2 \sigma_S^2/(\sigma_D^2+ \sigma_S^2)$. Eq. (\ref{P2}) along with the definitions given in Eqs. (\ref{Phi2},\ref{X}) can be used to quantify the oscillations as well as decoherence for the neutrinos with Gaussian wave packets at the source and detector and travelling in the weak gravity regime.

Several interesting aspects of Eqs. (\ref{P2},\ref{Phi2},\ref{X}) can be discussed at this stage.
\begin{itemize}
\item The oscillation phase obtained in Eq.(\ref{Phi2}) is, in general, different from the one obtained assuming neutrinos as plane waves. In the case of the latter, $\Phi^{mn}_{ij} = \Phi_i^{m} - \Phi_j^{n}$ \cite{Swami:2020qdi}. The difference becomes negligible if the neutrino wave packets at production and detection follow $\vec{p}^D = \vec{p}^S$.
\item By definition all the damping factors are non-negative, i.e. ${\bf X}_{ij}^{mn} \ge 0$. Further, they are symmetric under the operation $(i, m) \rightleftharpoons (j,n)$.
\item As the neutrinos move along their trajectories, all ${\bf X}_{ij}^{mn}$ increase because of their dependence on the travelled distance. Clearly, the smaller a particular ${\bf X}_{ij}^{mn}$ is, the later in time the corresponding exponential term will be decaying in Eq. (\ref{P2}). However, a particular ${\bf X}_{\hat{i} \hat{i}}^{\hat{m} \hat{n}}$ can be chosen, as the one with the smallest magnitude among all ${\bf X}_{ii}^{mn}$, and it is easy to see that the probability expression, Eq. (\ref{P2}), does not depend on ${\bf X}_{\hat{i} \hat{i}}^{\hat{m} \hat{n}}$. Note that to utilize this freedom to choose the index $\hat{i}$ effectively, we require that the corresponding $U_{\alpha \hat{i}}$ are non-vanishing. Otherwise, $P_{\alpha \beta}$ is already independent of ${\bf X}_{\hat{i} \hat{i}}^{\hat{m} \hat{n}}$. The effective damping factor can, therefore, be parametrized as
\be \label{D}
D_{ij}^{mn} = {\bf X}_{ij}^{mn} - {\bf X}_{\hat{i} \hat{i}}^{\hat{m} \hat{n}}\,,
\ee
such that ${\bf X}_{\hat{i} \hat{i}}^{\hat{m} \hat{n}}$ is the smallest among ${\bf X}_{ij}^{mn}$ and corresponding $U_{\alpha \hat{i}} \neq 0$. The same expression of probability holds with ${\bf X}_{ij}^{mn}$ are now replaced by $D_{ij}^{mn}$ in Eq. (\ref{P2}).
\item For decoherence, it is necessary that $ \sigma_D,\sigma_S\neq 0$. If any of the two vanish, $\bar{\sigma}$ vanishes too. Since $\sigma_{D,S}\rightarrow 0$ also marks the perfect production or detection mechanism, a precision information of the production process of neutrinos at the source end or precision in identification of a transition at the detector end allows less room for decoherence.
\item When neutrino travels large enough distance, one eventually finds all $e^{- D_{ij}^{mn}} \to 0$, except for $i=j=\hat{i}$ and $m=\hat{m}$, $n=\hat{n}$ for which $e^{- D_{\hat{i} \hat{i}}^{\hat{m} \hat{n}}} = 1$. In this limit, one finds 
\be \label{P_saturated}
P_{\alpha \beta} \to |U_{\beta \hat{i}}|^2\,, \ee
provided $U_{\alpha \hat{i}} \neq 0$ as discussed earlier. The value of saturated probability, therefore, depends only on the neutrino mixing parameters. 
\end{itemize}
In order to illustrate the last point with more clarity, consider a three flavor case with $U_{\alpha 1} \neq 0$. As the neutrinos travel more the parameter $D_{ij}^{mn}$ keep increasing. The largest  $D_{ij}^{mn}$ will cross some specified value of irrelevance (say $1$)  first, will lead its corresponding term to insignificant values the earliest with increasing distance. Gradually various terms will keep dropping off as neutrinos move forward from $S$ to $D$. Eq. \eqref{P2} can be written in the form 
 \begin{equation} \label{MassBreak}
P_{\alpha\beta}=\underbrace{\frac{ \sum \limits_{i,m,n} U_{\beta i}^{*}U_{\alpha i} U_{\beta i}U_{\alpha i}^{*}  e^{- \iota	\Phi_{ii}^{mn} -D^{mn}_{ii}}}{{\sum \limits_{i,m,n} U_{\alpha i} U_{\alpha i}^{*}  e^{- \iota	\Phi_{ii}^{mn} -D^{mn}_{ii}}} }}_{\textbf{I}} + \underbrace{\frac{\sum \limits_{i\neq j,m,n} U_{\beta i}^{*}U_{\alpha i} U_{\beta j}U_{\alpha j}^{*}  e^{- \iota	\Phi_{ij}^{mn} -D^{mn}_{ij}}}{\sum \limits_{i,m,n} U_{\alpha i} U_{\alpha i}^{*}  e^{- \iota	\Phi_{ii}^{mn} -D^{mn}_{ii}}}}_{\textbf{II}},
\end{equation}
such that the terms are arranged as : \textbf{I } which  involves the same mass on various path interference and  \textbf{II} which  involves different mass species on various path interference. Given the hierarchical structure set up, the term  \textbf{II} starts with $D_{12}^{11}$ (which is the smallest in   \textbf{II}). Therefore, all other $D_{ij}^{mn}$  carrying exponentials will decay before the decay of the first term in  \textbf{II}. So at the stage when $ D_{12}^{11}$ carrying exponential is the only significant term in \textbf{II}, all other terms in \textbf{I} apart from smallest mass term $i=1$, also become irrelevant (as they are all larger than  $ D_{12}^{11}$ ).  Thus, in an $n$ flavour case, the last few non-trivial relevant terms in the probability of transitions after travelling sufficiently far from the source are those with $D_{11}^{mn}$ and $D_{12}^{11}$. Now interestingly, as soon as the last remaining exponential in \textbf{II} turns insignificant, the probability of transition already saturates, as when \textbf{II} $\rightarrow 0$, we have
\begin{equation} \label{53}
P_{\alpha\beta}  \rightarrow \textbf{I}\longrightarrow \frac{ \sum \limits_{m,n} U_{\beta 1}^{*}U_{\alpha 1} U_{\beta 1}U_{\alpha 1}^{*}  e^{- \iota	\Phi_{11}^{mn} -D_{11}^{mn}}}{\sum \limits_{m,n} U_{\alpha 1} U_{\alpha 1}^{*}  e^{- \iota	\Phi_{11}^{mn} - D_{11}^{mn}}} = U_{\beta 1}^{*} U_{\beta 1}= |U_{\beta 1}|^2 .
\end{equation}
 Therefore, the decoherence is decided by the decay of same (larger) path interference term between two smallest mass species wave packets,  i.e. through $D_{12}^{11} \rightarrow 1$.
 In the case when the $U_{\alpha i} =0$ for all $i < \hat{i}$ the decoherence condition gets modified to  $D_{\hat{i},\hat{i}+1}^{11} \rightarrow 1$.

 As it  may be apparent, the number of flavours had no explicit role to play the condition for decoherence  is general and applicable to neutrinos travelling in flat as well as curved spacetime as far as gravity is weak. It also holds for $n$ number of flavours and multiple classical path neutrinos may take to reach to the detector from a given source. Thus, it will work for cases where we have only one path connecting the detector and the source (as in case of radial propagation or for particle with non-zero angular momentum when the source and detector are on the same side) as well as multi-path consideration (non-radial propagation, i.e., lensing).


\section{Decoherence in the Schwarzschild metric} \label{SchwarzschildDecoherence}
We now discuss the decoherence in the presence of Schwarzschild background in order to quantify the effects that arise due to the curvature of spacetime. The Schwarzschild metric quantifying the gravitational field of a spherically symmetric body is written as 
\be \label{s_metric}
ds^2= B(r)\, dt^2 - B^{-1}(r)\, dr^2 - r^2 d\theta^2 - r^2\, \sin^2\theta\, d\phi^2\,,
\ee
where $B(r) = \left(1- R_S/r \right)$ and $R_S$ is Schwarzschild radius. As is the general practise, neutrinos are assumed to travel on null geodesics of this metric. The spherical symmetry of the system confines these geodesics on a plane which can be chosen as $\theta = \pi/2$ without loss of generality.  The phase $\Phi^m_i$, defined in Eq. (\ref{Phi}), can then be evaluated for classical trajectories between the source and detector. The justification for considering such classical trajectories and details of evaluation of the phase have been described in detail in our previous work \cite{Swami:2020qdi}. The evaluation of phase depends on two qualitatively different cases corresponding to radial and non-radial trajectories. In the radial case, there is only one trajectory available for the neutrinos, whereas for the non-radial case the number of trajectories (in a plane of constant $\theta$) may be 1 or 2 depending upon whether the source and the detector are on the same side of Schwarzschild mass or different respectively. We will discuss all such cases now.

\subsection{Radial propagation}
In order to remain in the regime of weak gravity limit, one has to consider both the source and detector on the same side of the Schwarzschild body. There exists only one classical trajectory for neutrinos in this case and therefore we drop the path indices from the phase and other relevant quantities. The evaluation of phase in this case gives \cite{Fornengo:1996ef}
\be \label{Phi_radial}
\Phi_j = \int_{r_S}^{r_D} \left(E_j \left( \frac{dt}{dr}\right) - p_j \right)\, dr \simeq \pm \frac{m_j^2}{2 E_0}\, (r_D - r_S)\,
\ee
at the leading order, where $r_D$ and $r_S$ are radial coordinate distance defined in the Schwarzschild coordinate system. In the evaluation of the above, we have used $dt/dr = \pm 1/B(r)$ for null trajectories. Also, $E_j$ and $p_j$ are constants of motion and are related by
\be \label{p_radial}
p_j(r) = \pm \frac{1}{B(r)}\, \sqrt{E_j^2 - B(r) m_j^2} \simeq \pm \frac{1}{B(r)}\, \left(E_j - B(r) \frac{m_j^2}{2 E_0}\right)\,.
\ee
The positive (negative) sign in the above expressions stands for neutrino travelling outward (inward). $E_0$ is the energy as measured by an observer at the infinity and it is constant along the null trajectory \cite{Fornengo:1996ef}.  Taking the momentum distribution defined at the source location (i.e. the maxima of the distribution as well as ${\bf X}_{ij}$ is defined at the location of the source) and following the definitions Eqs.(\ref{Phi_sub},\ref{X}) along with Eqs.(\ref{Phi_radial},\ref{p_radial}), we evaluate the decay factor as 
\be \label{X_radial}
{\bf X}_{ij} \simeq \bar{\sigma}^2 \frac{m_i^4 + m_j^4}{8 E_0^4}\,B(r_S)\, |r_D - r_S|^2\, =\, \bar{\sigma}^2 \frac{m_i^4 + m_j^4}{8 E_{\rm loc}^4 B(r_S) }\, |r_D - r_S|^2\,, \ee
at the leading order in $m_i/E_0$, see Appendix \ref{Tetrad}. Here,
\be \label{}
E_{\rm loc} \equiv E_{\rm loc}(r_S)  = \frac{E_0}{\sqrt{B(r_S)}}\,,
\ee
is the energy of neutrinos (in the equal energy approximation) as measured by a local observed situated at the source. Identifying the lightest mass eigenstate as $m_1$ and the second lightest as $m_2$, it is straight-forward to see from Eq.(\ref{D}) that the smallest non-zero $D_{ij}$ corresponds to
\be \label{D_radial}
D_{12} \simeq \bar{\sigma}^2 \frac{m_2^4 - m_1^4}{8 E_{\rm loc}^4 B(r_S)}\, |r_D-r_S|^2\,.
\ee
The decoherence distance, i.e. the distance at which the oscillation probability gets depleted  by  atleast a factor of $e^{-1}$, is then quantified by setting $D_{12} = 1$. In other words, the neutrinos will decoher while travelling radially inward or outward if
\be \label{declength_radial}
\frac{|r_D-r_S|}{\sqrt{B(r_S)}} \ge 2 \sqrt{2} \frac{E_{\rm loc}^2}{ \bar{\sigma} \sqrt{m_2^4 - m_1^4}}.
\ee
For given $E_{\rm loc}$, $m_1$, $m_2$, $\bar{\sigma}$ and $r_S$, one can obtain the location $r_D$ where the decoherence will set in  Equivalently, one can infer about the absolute neutrino mass scale from decoherence length if the other parameters and squared difference of masses are known.

In the derivation of Eq. (\ref{declength_radial}), we have used a source wave function, Eq. (\ref{psiS}), expanded in terms of momentum distribution function as seen by an observer located at the source. One can also perform similar analysis in terms of a momentum distribution function  specified for an observer at infinity. Assuming again a Gaussian distribution in this case, the condition equivalent to Eq. (\ref{declength_radial}) is obtained as
\be \label{declength_radial_asymptotic}
|r_D-r_S| \ge 2 \sqrt{2} \frac{E_0^2}{ \bar{\sigma} \sqrt{m_2^4 - m_1^4}}\,. \ee
The above differs from Eq. (\ref{declength_radial}) by an extra  factor of $\sqrt{B(r_s)}$. The decoherence as perceived by different observers is not identical in curved spacetime as the momentum distributions are defined differently in different frames. The result in Eq. (\ref{declength_radial_asymptotic}) is in a qualitative agreement with the ones derived in \cite{Chatelain:2019nkf, Petruzziello:2020wea}. However, we get a different combination of neutrino masses in Eq. (\ref{declength_radial_asymptotic}) in comparison to the results obtained in \cite{Chatelain:2019nkf, Petruzziello:2020wea}.

A few important points can be noted in the context of the above results. The energies appearing in the oscillation phase and damping factor are different in general. Moreover, as can be seen from Eq. (\ref{declength_radial_asymptotic}), the decoherence coordinate $r_D$ is insensitive to the Schwarschild parameter $R_S$ at the leading order in $m_j/E_{\rm loc}$ from a perspective of an asymptotic observer. Note that it is the only radial coordinate determination which is independent of $R_S$ in this case. The physical spatial distance neutrinos travel radially before their wave packet separates  depends on $R_S$. Such distance can be obtained as 
\be
L_{p} =\int_{r_S}^{r_D} \frac{1}{\sqrt{B(r)}}\, dr \simeq r_D- r_S + \frac{R_S}{2}\, \ln{\left(\frac{r_D}{r_S}\right)}\,.\ee
Consequently, the spatial distance $L_p$ travelled by the neutrino turns out to be greater that that in the Schwarzschild background. Hence, the coherence is maintained for relatively greater spatial distance in curved geometry.

\subsection{Non- Radial propagation with a single trajectory}
This case corresponds to situation that the source and the detector are on the same side of the gravitating mass  with $r_S<r_D$ and the neutrinos are created with non-zero angular momentum.  In this case, the proper time taken in moving from $r_S$ to $r_D$
\begin{equation} 
\tau_i \approx \int_S^D dr \frac{m}{E_0}\left(1+\frac{m^2}{2E_0^2}+\frac{L^2}{2E_0^2r^2}-\frac{r_S}{r}\left(\frac{m^2}{2E_0^2}+\frac{L^2}{2E_0^2r^2}\right)\right).
\end{equation}
Now using $L= E_0 b v_\infty= E_0b(1-\frac{m^2}{2E_0^2})$,  upto first order in $m/E_0$  we get,
\begin{equation} \label{NRadSameSide}
\tau_i =  \frac{m}{E_0}\left((r_D-r_S)+\frac{b^2}{2}\left(\frac{1}{r_S}-\frac{1}{r_D}\right)\right)-r_S \frac{m}{E_0}\left(\frac{b^2}{4}\left(\frac{1}{r_S^2}-\frac{1}{r_D^2}\right)\right).
\end{equation}
Further, in case of single path ($m=1$),  the exponents $D_{ij}^{11}$ can be evaluated for the Schwarzschild geometrical background in the weak gravity case as
\begin{equation} \label{ExponentD}
	D_{ij}^{11}= \boldsymbol{{X}}_{ij}^{11} -\boldsymbol{{X}}^{11}_{11}  = \frac{\bar{\sigma}^2}{2}\left(|\vec{X}_i|^2-| \vec{X}_1|^2+ | \vec{X}_j|^2-| \vec{X}_1|^2\right)
 \end{equation}
  with
\begin{equation} \label{XinLens}
	|\vec{X}_i|^2 \equiv	\frac{m_i^4 B(r_S)}{4 E^4_0} R^2 \left( 1- \frac{b^2}{2r_S r_D} + \frac{r_S}{R}\right)^2 \approx \frac{m_i^4 B(r_S)}{4 E^4_0} R^2 \left( 1- \frac{b^2}{r_S r_D} + \frac{2r_S}{R}\right).
\end{equation}

Since, $\vec{X} ^m_i = \partial_{\vec{p}}\phi_i |_{\vec{p}_i^S}$, for the Gaussian wave packet we have
\bea
\boldsymbol{X}^{mn}_{ij}=  \frac{\tilde{\sigma}^2}{8g^{00} E_0^2}\left((m_i\tau_i)^2+ (m_j\tau_j)^2\right),
\eea
 since $|\vec{X}_i^m|^2 \equiv \sum \limits_{a=1}^{3}X_i^{m,a} X_{i,a}^m = (m_i\tau_i^m)^2/4g^{00}$, 
where $X_i^{m,a}$ is the projection of $ \vec{X}_i^m$ in terms of local tetrad basis. Thus, the decoherence controling parameter
\bea
D_{12}^{11}= \frac{\tilde{\sigma}^2 B(r_S)}{8 E_0^2}[(m_2 \tau_2)^2- (m_1\tau_1)^2 ]. \label{DecohParamSchwarzshild}
\eea

In order to cause appreciable decoherence one has to attain particular $D_{12}^{11} \rightarrow 1$ value. It can be shown that in order to traverse a particular amount of proper time, one has to travel more in terms of physical distance in the Schwarzschild spacetime compared to the flat space.  Therefore, it follows that for non-zero $b_1$ (non-radial case)  one has to travel  more in radial co-ordinate equivalently lapsing more  physical distance spatially, see Appendix \ref{App2} . 

\section{Lensing} \label{Lensing}
We now discuss the non-radial propagation with the source and detector located on the opposite sides of gravitating object which is essentially the case for neutrino lensing phenomena. The geometrical configuration of this case and derivation of phase $\Phi^m_i$ are discussed in our previous work \cite{Swami:2020qdi} in detail. It is evaluated as
\be \label{}
\Phi^m_i = \int_{r_S}^{r_D} \left( E_i \left(\frac{dt}{dr}\right) - p_i - J_i \left(\frac{d\phi}{dr}\right) \right)\, dr \simeq  \frac{m_i^2}{2E_0}(r_S + r_D)\left( 1- \frac{b_m^2}{2r_S r_D} + \frac{R_S}{r_S + r_D}\right)\,, \ee
where the angular momentum $J_i$ has been conveniently parametrized in terms of impact parameter $b_m$. The second equality in the above equation is obtained in the weak gravity limit $r_{S,D} \gg R_S$ as well as  $r_{S,D} \gg b_m$. A straight-forward evaluation of $\vec{X}^m_i$ then gives
\be \label{Xim2}
\left|\vec{X}_i^m\right|^2  \simeq	\frac{m_i^4}{4 E_{\rm loc}^4 B(r_S)} (r_S + r_D)^2 \left( 1- \frac{b_m^2}{2r_S r_D} + \frac{R_S}{r_S + r_D}\right)^2 \approx \frac{m_i^4}{4  E_{\rm loc}^4 B(r_S)}  (r_S+r_D)^2 \left( 1- \frac{b_m^2}{r_S r_D} + \frac{2R_S}{r_S + r_D}\right)\,
\ee
at the leading order in $m_i/E_{\rm loc}$. 

Let us now quantify the decoherence in terms of the effective damping factor $D_{ij}^{mn}$. Given source and detector on the opposite sides of the gravitating object, there are two classical trajectories on which neutrinos can travel in this case. These trajectories are distinguished by their impact factor $b_1$ and $b_2$. Identifying $x$-axis with the line connecting neutrino source and Schwarzschild body, one can choose the impact parameters such that $b_1 \le b_2$ for $y \ge 0$. Further, we can arrange neutrino masses such that $m_1 < m_2 < ...<m_n$. Therefore,  an appropriate damping factor, as defined in Eq. (\ref{D}), for $y \ge 0$ is determined as
\be \label{D_lensing}
D_{ij}^{mn} = {\bf X}_{ij}^{mn} - {\bf X}_{11}^{11} \approx \frac{\bar{\sigma}^2 (r_S + r_D)^2}{8 E_{\rm loc}^4 B(r_S)} \left(1+\frac{2 R_S}{r_S+r_D}\right) \left[m_i^4 \left( 1 - \frac{b_m^2}{r_S r_D} \right) + m_j^4 \left( 1 - \frac{b_n^2}{r_S r_D} \right) - 2 m_1^4 \left( 1 - \frac{b_1^2}{r_S r_D} \right) \right]\,.
\ee
It can be seen that decoherence can arise in two qualitatively different ways: (a) due to mass difference between the lightest and the second lightest neutrino mass eigenstate, i.e. when $i$ or $j \neq 1$ and, (b) because of path difference even when $i=j=1$. Clearly, the second effect is negligible as it arises at sub-leading order. It is noteworthy that the contribution that arise through (b) actually decreases (recall that $b_1 < b_2$ in the region of our interest) the effective damping factor and therefore implies relatively increased length of coherence. However, in case of weak lensing these modifications are extremely tiny and seem irrelevant from the practical point of view. 

Even in the non-radial propagation case, the decoherence is dominantly governed by mass difference between the lightest and the second lightest neutrino and, therefore, the relevant damping factor is $D_{12}^{11}$. We find that the coordinate distance at which the lensing probabilities get diluted by at least a factor of $e^{-1}$ is given by a condition
\be \label{declength_lensing}
\frac{(r_D + r_S)}{\sqrt{B(r_S)}}\left(1 -\frac{b_1^2}{2 r_S r_D} + \frac{R_S}{r_D + r_S}\right) \ge 2 \sqrt{2} \frac{E_{\rm loc}^2}{ \bar{\sigma} \sqrt{m_2^4 - m_1^4}}\,.
\ee
Eq. (\ref{declength_lensing}) can readily be applied to the neutrino lensing case in order to estimate the distance till which the coherent oscillations will last. 

For a momentum distribution function specified in the frame of an asymptotic observer, the condition equivalent to Eq. (\ref{declength_lensing}) is obtained as
\be \label{declength_lensing_asymptotic}
(r_D + r_S) \left(1 -\frac{b_1^2}{2 r_S r_D} + \frac{R_S}{r_D + r_S}\right) \ge 2 \sqrt{2} \frac{E_0^2}{ \bar{\sigma} \sqrt{m_2^4 - m_1^4}}\,.\ee
Unlike the decoherence condition obtained in the radial case, Eq. (\ref{declength_radial_asymptotic}), the above explicitly depends on the Schwarzschild mass. A naive estimation of decoherence length was already given in our previous work, see Eq. (32) in \cite{Swami:2020qdi}. The result we obtain here through a more careful and explicit treatment is in a qualitative agreement. However, there is an important difference. The decoherence length given in Eqs. (\ref{declength_lensing},\ref{declength_lensing_asymptotic}) depends not only on the difference of squared neutrino masses but also on the sum of them. Apart from this, there is also a difference of factor $2$ between the two results.

\section{Phenomenological Implications}
\label{pheno}
The main result Eqs. (\ref{declength_lensing},\ref{declength_lensing_asymptotic}), obtained assuming neutrinos as Gaussian wave packets, reveals some phenomenologically useful aspects of lensing which complements our previous study of the same but with neutrinos as plane waves \cite{Swami:2020qdi}. The noteworthy features are the following.
\begin{itemize}
\item In comparison to the plane wave approach, the neutrino lensing probabilities in the present case eventually saturate to some particular values and the oscillation seizes. The distance at which these effects become sizeable is given by Eqs. (\ref{declength_lensing},\ref{declength_lensing_asymptotic}). This distance depends not only on the energy and the width of neutrino wave packets but also on absolute neutrino mass scale. For example, for fixed $\bar{\sigma}$, $E_0$ and $\Delta m_{21}^2 = m_2^2 - m_1^2$, neutrinos maintain coherent oscillation for relatively longer distance if they are hierarchical (i.e. $m_1 \ll m_2$). For less hierarchical neutrinos (i.e. $m_1 \simeq m_2$), the decoherence occur at relatively shorter distance. This feature is not only restricted to lensing phenomena but also holds for radial propagation as it can be seen from Eqs. (\ref{declength_radial},\ref{declength_radial_asymptotic}). 
\item Distances within which the coherent oscillations occur, the dependency of flavour transition probability on absolute neutrino mass scale arises only through path differences between neutrino trajectories as discussed in \cite{Swami:2020qdi}. Hence, lensing is essential in that case. 
\item For the neutrino lensing, depending on the locations of source, gravitating object and detector and for given energy and widths of neutrino wave packets, the system can be found in coherent or decoherent regime. One sees qualitatively very different pattern of neutrino lensing probabilities in these regimes as can be seen from our present and previous studies \cite{Swami:2020qdi}.
\end{itemize}

\begin{figure}[ht!]
\centering
\subfigure{\includegraphics[width=0.43\textwidth]{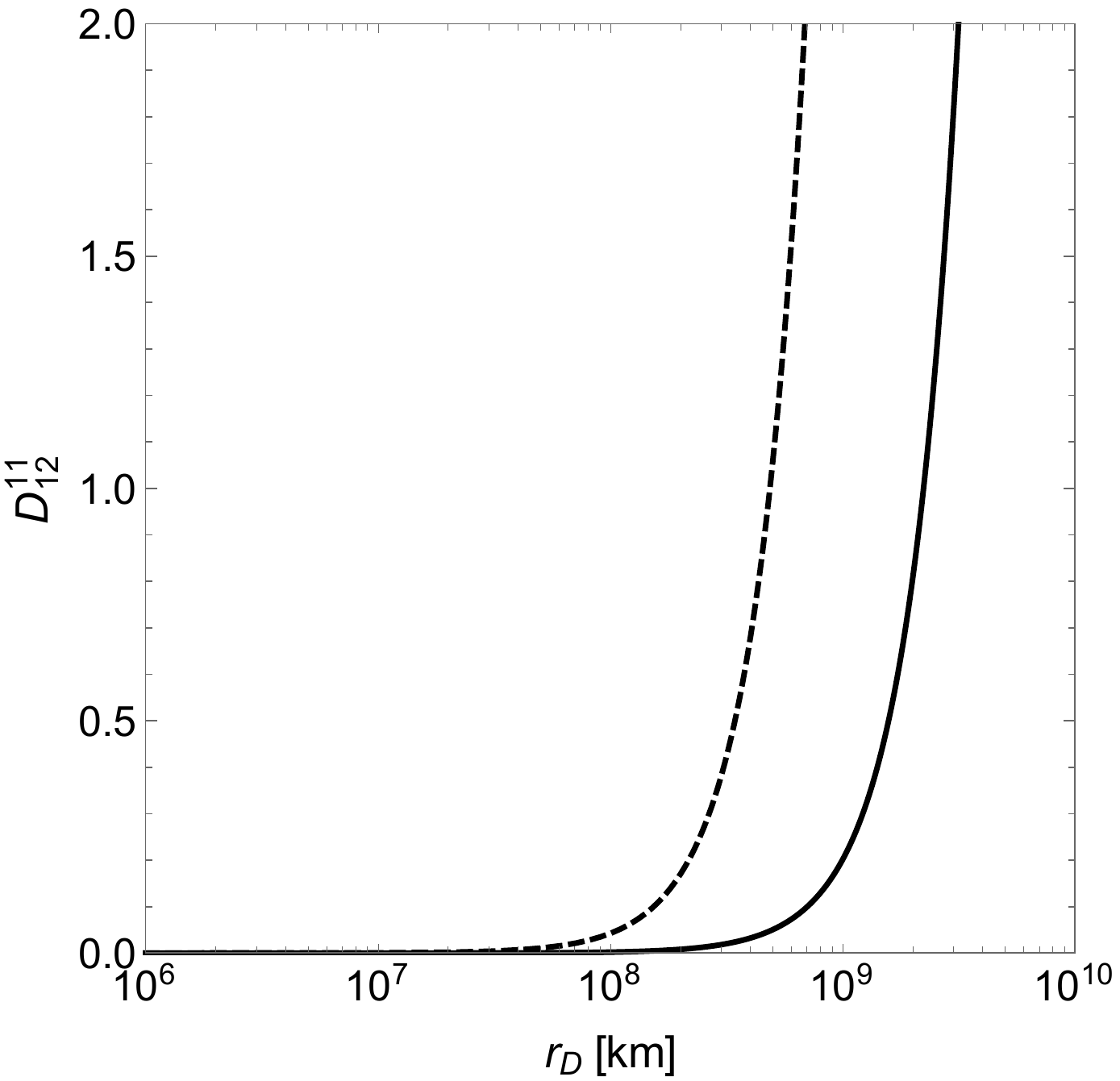}}\hspace*{0.1cm}
\subfigure{\includegraphics[width=0.45\textwidth]{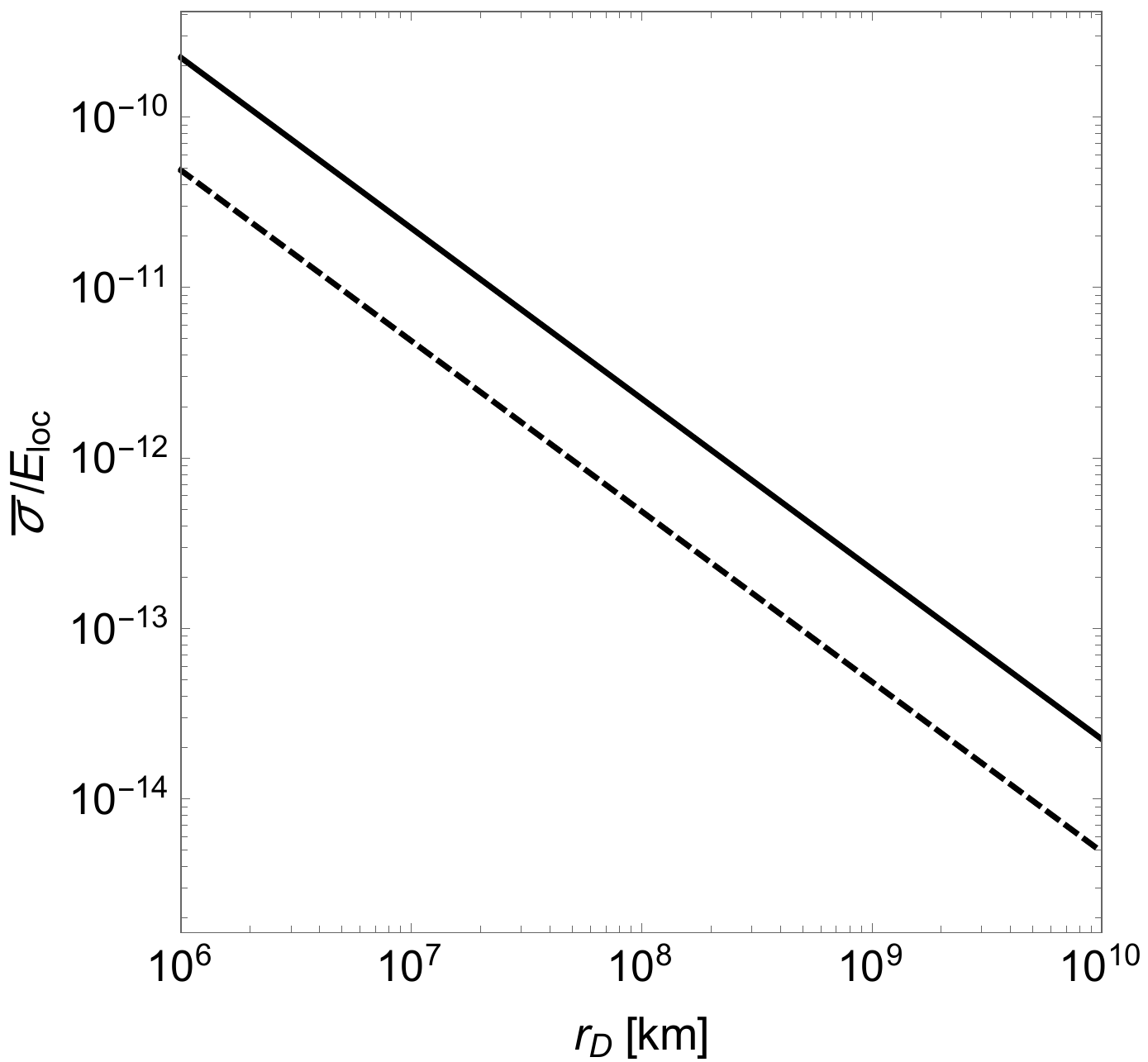}}
\caption{Left panel: the damping factor $D_{12}^{11}$ as function of $r_D$ for $\bar{\sigma}/E_{\rm loc} = 10^{-13}$. Right panel: contours corresponding to $D_{12}^{11} = 1$. In both the panels, the solid (dashed) line corresponds to $m_1 = 0$ ($m_1 = 0.1$) eV. The other parameters are $r_S = 10^5 r_D$, $R_S = 3$ km, $m_2^2 -m_1^2 = 10^{-3}\,{\rm eV}^2$ and $E_{\rm loc} = 10$ MeV.}
\label{fig1}
\end{figure}

To make the above points more clear, we now estimate the decoherence length for an example of Sun-Earth based lensing system discussed earlier in detail in \cite{Swami:2020qdi} using the condition, Eq. (\ref{declength_lensing}). We consider $R_S = 3$ km, $E_{\rm loc} = 10$ MeV and $r_S = 10^5 r_D$ as taken earlier. For simplicity, we consider collinear case in which the source of neutrino, the detector and gravitating body lie on the same line. As discussed before, deviation from this alignment does not lead to significantly different results for decoherence. We then compute the damping factor of interest, $D_{12}^{11}$, for given wave packet width in the momentum space and for different values of detector location. At $r_D$ where $D_{12}^{11} = 1$, the deviation from the saturation value of the transition probability is damped by a factor of $1/e$. For $D_{12}^{11} = n$, this deviation further weakens roughly by a factor of $1/e^n$. We also compute correlations between $\bar{\sigma}$ and $r_D$ for a fixed $D_{12}^{11}$ value, which we choose to be unity (one can take any reference value, with larger value indicating more effective saturation). The estimation is done for two different values of the lightest neutrino mass $m_1$ but keeping $m_2^2-m_1^2 = 10^{-3}\, {\rm eV}^2$ fixed. The results are displayed in Fig. \ref{fig1}.  It can be seen that for a given finite width of the wave packet, decoherence occurs relatively at larger distance for hierarchical neutrino masses.

\begin{figure}[ht!]
\centering
\subfigure{\includegraphics[width=1\textwidth]{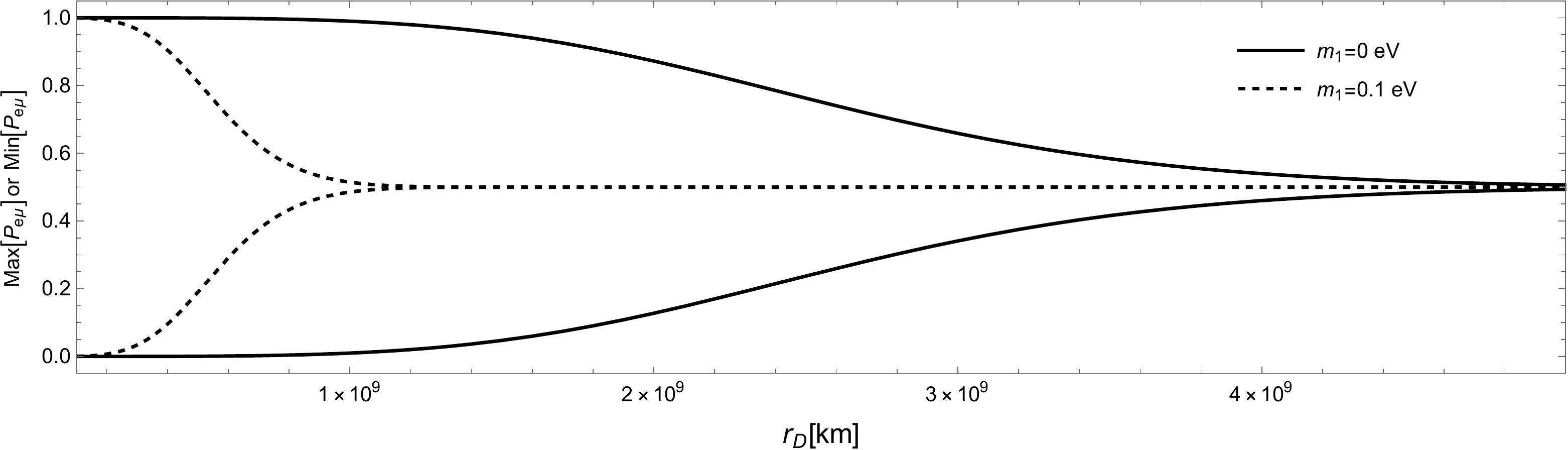}}\hspace*{0.1cm}
\caption{Maximum and minimum transition probability envelop as a function of $r_D$ for two flavour case.  The solid (dashed) line corresponds to $m_1 = 0$ ($m_1 = 0.1$) eV. The mixing angle is $\alpha=\pi/4$ and all the other parameters are as given in the caption of Fig. \ref{fig1}.}
\label{fig2}
\end{figure}

As discussed above, as $D_{12}^{11}$ increases the transition  probability more and more effectively saturates to a value determined by the mixing angle $\alpha$. To demonstrate this effect more clearly, we compute the transition probability $P_{\alpha \beta}$ in the two flavour case (with $\alpha, \beta$ marking either of electron $(e)$ or muon type $(\mu)$ neutrino flavor), taking the mixing angle $\alpha = \pi/4$ and show the dependency of $P_{\alpha \beta}$ on $r_D$ near the decoherence distance. This is displayed in Fig. \ref{fig2}. Note that the probability oscillates very rapidly at large distance nearby the decoherence length. Thus to suppress these effects, we plot the probability envelop made from maximum and minimum of the probability distribution rather than the value of probability itself. To generate this envelop, we select the maxima and the minima of probability over a certain range $\Delta r_D$ nearby particular value of $r_D$ and associate these values to averaged $r_D$. For the given plot $r_D$ ranges from $10^8$ km to $  5\times 10^9$ km and we choose $\Delta r_D =2.45 \times 10^6$ km for determining the maxima and the minima of the transition probability. This window  corresponds to a total of 4000 data points on the plot. The minimum and  maximum values of the transition probability are then used to generate the envelopes displayed in  Fig. \ref{fig2}.  


It can be seen that amplitude of the probability gradually decreases and the probability settles to a value determined by mixing matrix element which in this case is $P_{e\mu} \rightarrow \sin^2 \alpha = 1/2$. From Fig. \ref{fig1} and \ref{fig2} we see the probability saturates at a radial distance $r_D$ where the  decoherence factor $D_{12}^{11}$ becomes greater or equal to 2 for both the cases $m_1=0$ eV and $m_1=0.1$ eV. We see in the $m_1 =0.1 $ eV case neutrino achieves probability saturation faster as compared with $m_1 =0$ eV case.   This demonstrates neutrino decoherence sensitivity to the absolute neutrino masses as discussed before. Beyond the decoherence length, the interference caused by lensing of neutrinos gets diminished and the probability saturates to the value as discussed in Eq. (\ref{P_saturated}).

\section{Summary}\label{Summary}
Gravitational lensing of neutrinos can reveal some interesting features of neutrino flavour oscillations which cannot be seen in usual oscillations in flat spacetime. For example, it has been shown that the transition probability obtained through lensing depends not only on the squared mass difference of the neutrinos but it is also sensitive to the absolute neutrino masses \cite{Swami:2020qdi}. Since lensing involves propagation of neutrinos over huge distances, a realistic study of this phenomena must include understanding of decoherence in the presence of gravitational background, which we carry out in this paper. Assuming neutrino wave functions as Gaussian wave packets of finite width for both the source and the detector, we first derive a general expression of transition probability in the wave packet formalism. It is seen that the wave packet approach not only gives rise to decoherence  but also modifies the oscillation phase if the mean values of momentum involved in the production and detection mechanisms are different. Interestingly, the efficiency of decoherence depends crucially not only on the detector variance in the distribution but also on that of the source.

We apply this general treatment to the radial and non-radial propagations of neutrinos in the background of Schwarzschild geometry. In a general spacetime, the amount of decoherence a wave packet suffers, gets decided by the proper time spent by an observer co-travelling with its maxima while propagating between the source and the detector location. Since for a given spatial distance between two points on a spatial hypersurface, the proper time elapsed along a geodesic connecting them is shorter in the Schwarzschild background, somewhat counter intuitively the wave packets have to travel more (spatially) in presence of gravity when compared to the flat spacetime, in order to achieve the same level of decoherence. 

Further, the non-radial propagation is studied viz-a-viz the gravitational lensing phenomenon. It is seen that the separation of neutrino wave packets in case of lensing depends on both the Schwarzschild mass and the classical path taken between the source and detector although at sub-leading order, see Eq. (\ref{declength_lensing}). It is also seen that the decoherence lengths in both the cases are sensitive to absolute neutrino mass scale through explicit dependence on both the sum and difference of the squared masses. Therefore, observing gravitational effects on neutrino oscillations even with presence of decoherence effect remains a viable avenue for obtaining mass hierarchy information of neutrinos. A realistic study of neutrino fluxes from astrophysical sources can be used to estimate the precision required in astrophysical or ground based neutrino observations to reveal such aspects.

\acknowledgments
HS would like to thank Council of Scientific \& Industrial Research (CSIR), India for the financial support through research fellowship award no. 514856. Research of KL is partially supported by the Department of Science and Technology (DST) of the Government of India through a research grant under INSPIRE Faculty Award (DST/INSPIRE/04/2016/000571). The work of KMP is partially supported by a research grant under INSPIRE Faculty Award (DST/INSPIRE/04/2015/000508) from the DST, Government of India.  KL is also grateful towards the hospitality of Physical Research Laboratory,  Ahmedabad, where part of this work was carried out.

\appendix

\section{Decoherence parameter} \label{Tetrad}
 Since the wave packet is assumed to have a sharp distribution around $\vec{p}^S$, we include upto first order expansion around it.
\bea
\Phi_i^m(\vec{p})=\Phi_i^m(\vec{p}^S) + (\vec{p}-\vec{p}^S)\cdot \nabla\Phi_i^m(\vec{p}^S).
\eea
We define $\vec{X}_i^m = \nabla\Phi_i^m(\vec{p}^S)$ as a gradient of the phase w.r.t. momentum defined at location $r$.  Further, along path $m$ the expression of phase
\bea
\Phi_i^m(\vec{p})=\int (p_i)_{\mu}dx^{\mu}
\eea
For a radial trajectory (with a diagonal metric) $p_{\mu}=(p_0,p_r,0,0)$. Further, for a timelike Killing vector $k^{\mu} =(1,0,0,0)$, $p_{\mu}k^{\mu}$ is a conserved quantity along the geodesic whose tangent is $p^{\mu}$. For asympototic region this conserved quantity $p_{\mu}k^{\mu}\rightarrow p_0\equiv E_0$. Further since the vector $\vec{X}_i^m $ is obtained from the spacelike gradient of the phase, we go to the local Lorentz (tetrad) basis (just for the convenience of rectilinear co-ordinate system), defining $p_a=e_a^{\mu}p_{\mu}$. Further, owing to the diagonal metric structure we can select $e_0^{\mu} =\sqrt{\eta_{00}/g_{00}}\delta_0^{\mu}$, leading to $E_0=\pm \sqrt{g_{00}[ \sum_{j=1}^3(p_j)^2]/\eta_{00}}$. Using the null geodesic approximation, we write
\bea
\Phi_i^m(\vec{p})=\int [(p_i)_{0}dt + (p_i)_r dr]\approx  \frac{m_i^2}{2E_{0i}}R,
\eea
leading to
\bea
|\vec{X}_i^m|^2 =\sum_j(X_i^m)_j(X_i^m)_j=B(r) \frac{m_i^4}{4E_{0i}^4}R^2.
\eea
If the analysis is done w.r.t. the montum distribution defined at the source  location, then 
\bea
|\vec{X}_i^m|^2 = B(r_S) \frac{m_i^4}{4E_{0i}^4}R^2.
\eea

\section{Decoherance in flat spacetime vs Schwarzschild spacetime } \label{App2}

In this section, we compare the probability saturation rate between the flat spacetime and the Schwarzschild spacetime when the source and the detector located at a fixed proper distance apart. 
We do this by comparing decoherence factors of these spacetimes, using the following equation
\begin{equation} \label{94}
\Delta_{ijSF}^{mn} \equiv D_{ij}^{mnS}-D_{ij}^{mnF} = \boldsymbol{X}^{mn}_{ij}-\boldsymbol{X}^{11}_{11}
-(\boldsymbol{X}^{mnF}_{ij}-\boldsymbol{X}^{11F}_{11}
),	
\end{equation}
where $DF_{ij}^{mnS}$ and $DF_{ij}^{mnF}$  are the decoherence factor for the  Schwarzschild and flat spacetime respectively. Here $m.n$ in  $DF_{ij}^{mnF}$ indicates that we are choosing the same proper distances in the flat spacetime corresponding to the spatial path length taken by the neutrino in the Schwarzschild spacetime.
Note that $\Delta_{ijSF}^{mn} < 0$ will correspond to faster neutrino decoherence in the flat spacetime as compared to the Schwarzschild spacetime. 
 We can write Eq. (\ref{94}) explicitly as
\begin{equation}
 \Delta_{ijSF}^{mn} = \frac{\sigma_D^2\sigma_S^2}{ 8E_0^2(\sigma_D^2+\sigma_S^2 )}\Bigg( B\left(r\right)(m_i\tau_i^{m})^2 -(m_i\tau_i^{mF})^2+ B\left(r\right)(m_j\tau_j^{n})^2 -(m_j\tau_j^{nF})^2 + 
2B\left(r\right)(m_1\tau_1^{1})^2 -2(m_1\tau_1^{1F})^2\Bigg)	.
\end{equation}
Through some algebraic manipulations, the above expresion can be re-written as
\begin{eqnarray}  \label{95}
 \Delta_{ijSF}^{mn} = \frac{\tilde{\sigma}^2}{ 8E_0^2}\left(m_i^2 
\Delta \tau_{iSF}^{m}\sum \tau_{iSF}^{m}+ m_j^2\Delta \tau_{jSF}^{n}\sum \tau_{jSF}^{n}- \frac{R_s(m_i\tau_i^{m})^2}{r}- \frac{R_s(m_j\tau_j^{n})^2}{r} + 2m_1^2\Delta \tau_{1SF}^{1}\sum \tau_{1SF}^{1}- \frac{2R_s(m_1\tau_1^{1})^2}{r}\right),	\nonumber\\
\end{eqnarray}
where $\Delta \tau_{iSF}^{m} = \tau_i^{m} -\tau_i^{mF}$ and $\sum \tau_{iSF}^{m} = \tau_i^{m} +\tau_i^{mF}$. 
Now the proper time elapsed by the particle (having asymptotic energy $E_0$) in the flat spacetime $(d\tau_F)$ and in the Schwarzschild spacetime $(d\tau)$, after travelling a proper spatial distance $dl$, has the following relation 
\begin{equation} \label{77}
d\tau_F =  \frac{1}{\sqrt{B(r)}}d\tau +O\left(\frac{m^2}{E_0^2}\right).
\end{equation}
Ignoring $O\left(\frac{m^2}{E_0^2} \right)$, we get
\begin{equation} \label{78}
\frac{d\tau_F}{d\tau} \simeq  \frac{1}{\sqrt{B(r)}} >1,
\end{equation}
which under weak field limit can be written as 
\begin{equation} \label{79}
\int (d\tau_F-d\tau)= \tau_F-\tau =  \frac{R_s}{2}\int \frac{1}{r}d\tau >0.
\end{equation}
We see that irrespective of the path taken by the particle, proper time taken in the flat space is more.\\ 

Therefore,  $\Delta \tau_{iSF}^{m} $ is always negative because each term in Eq. \eqref{79} turns negative. Hence we see for the same spatial distance between the neutrino source and the detector, the neutrino transition probability in flat spacetime will saturate faster in comparison to the Schwarzschild spacetime.


%

\end{document}